\documentclass[aps,prl,preprint,showpacs,preprintnumbers,amsmath,amssymb,dvipdfm]{revtex4-1}
\usepackage[dvipdf]{graphicx}
\usepackage{dcolumn}
\usepackage{bm}
\usepackage[sort&compress]{natbib}

\newcommand{\erf}{\mathrm{erf}}

\begin{document}

\title{Mechanical signaling via nonlinear wavefront propagation in a mechanically-excitable medium}
\author{Timon Idema$^{1,2}$}
\author{Andrea J. Liu$^1$}
\affiliation{$^1$Department of Physics and Astronomy, University of Pennsylvania, Philadelphia, PA, USA\\ $^2$Department of Bionanoscience, Kavli Institute of Nanoscience, Delft University of Technology, Delft, The Netherlands}

\begin{abstract}
Models that invoke nonlinear wavefront propagation in a chemically excitable medium are rife in the biological literature. Indeed, the idea that wavefront propagation can serve as a signaling mechanism has often been invoked to explain synchronization of developmental processes. In this paper we suggest a new kind of signaling based not on diffusion of a chemical species but on the propagation of mechanical stress. We construct a theoretical approach to describe mechanical signaling as a nonlinear wavefront propagation problem and study its dependence on key variables such as the effective elasticity and damping of the medium.
\end{abstract}

\maketitle

The physical phenomenon of nonlinear wavefront propagation in an excitable medium is widely exploited by biological systems to transmit signals across many cells.  For example, when the slime mold {\it Dictyostelium} begins to aggregate to form a fruiting body, wavefronts of the molecule cAMP propagate across the amoeba colony~\cite{devreotes89,levine91,lee96}. Although cAMP itself spreads diffusively, wavefronts of cAMP propagate ballistically across the amoeba colony because the colony is chemically excitable: when the local concentration of cAMP exceeds a threshold, further local release of the species is triggered~\cite{levine91,lee96}. Similar wavefronts of calcium and potassium, respectively, signal fertilization in eggs~\cite{lechleiter92} and the onset of spreading cortical depression~\cite{gorelova83}, associated with migraine auras.   

In recent years, however, there has been a growing recognition that mechanics plays an important role in biology, and that many cells sense and respond not only to chemical stimuli but also to mechanical stimuli~\cite{altman02,discher05,vogel06,discher09,subramony13}. This raises the possibility that mechanosensing at the cellular level could give rise to collective phenomena at larger length scales such as collective cell migration~\cite{bois11,serra12}. 

In this paper, we consider the possibility of mechanically-induced waves at the tissue level. Biological systems are typically overdamped, so they do not support sound waves and stress cannot propagate ballistically. However, mechanical signaling via ballistic propagation of a nonlinear wavefront can occur in a mechanically-excitable medium, much as chemical signaling via ballistic propagation of a nonlinear wavefront can occur in a chemically-excitable medium. It has recently been suggested that two different biological systems might be mechanically-excitable:  the early {\it Drosophila} embryo~\cite{idema13plos} and the developing heart~\cite{majkut13}. The early {\it Drosophila} embryo supports mitotic wavefronts: nuclei at the poles of the embryo tend to divide first, giving rise to a mitotic wavefront separating dividing nuclei from those that have not yet divided.  This wavefront propagates across the entire embryo~\cite{foe83}. Likewise, the heart tube of the avian embryo beats via contractile wavefronts that are initiated at one end of the tissue and propagate across the heart tube with each beat~\cite{dejong}. These two examples are very different in biological details but share a key common feature--that the nuclei in the case of the {\it Drosophila} embryo and cardiomyocytes in the case of the heart tube both generate stresses as they proceed through mitosis and contraction, respectively.  It is therefore important to develop a general physical understanding of the basic properties needed to yield mechanically-induced nonlinear wavefronts.

Here we introduce two minimal models of mechanically-excitable media.  In each case, stress can be released at certain sites, or nodes, if the local stress exceeds some threshold value.  In the case of the early {\it Drosophila} embryo, these nodes would represent cell nuclei, while in the case of the developing heart tube, the nodes would represent cardiomyocytes (heart cells that can contract).  In our models, the released stress is transmitted through the damped elastic medium, potentially causing further release of stress at other nodes.   We solve these models and identify characteristic features exhibited by nonlinear wavefronts in such systems.

We start by considering two simple examples of media that can support mechanical stress and have overdamped mechanics. The elasticity is characterized by Lam\'e coefficients~$\lambda$ and~$\mu$, or equivalently by the Young's modulus~$E$ and dimensionless Poisson ratio~$\nu$ within linear elasticity theory~\cite{landauelasticity}. These parameters relate the stress $\sigma_{ij}$ inside the elastic material to its strain (deformation) $\varepsilon_{ij}$:
\begin{equation}
\label{stressstrain3D}
\sigma_{ij} = \frac{E}{1+\nu} \left[\varepsilon_{ij} + \frac{\nu}{1-(d-1)\nu} \varepsilon_{kk} \delta_{ij} \right],
\end{equation}
where $d$ is the number of dimensions (2 or 3), and summation over repeated indices is implied. The strain~$\varepsilon_{ij}$ is defined in therms of the displacement vector $u_i$ of the elastic material: $\varepsilon_{ij} = \frac12 (\partial_j u_i + \partial_i u_j)$. To avoid confusion we will follow the usual convention and label the two-dimensional versions of the parameters with a subscript 2, they are related to their three-dimensional counterparts by $E_2 = E / (1-\nu^2)$ and $\nu_2 = \nu / (1-\nu)$~\cite{landauelasticity}. The force per unit area is given by the divergence of the stress: $P_i = \partial_j \sigma_{ij}$.

In the simplest model that we consider, corresponding to a thin elastic film that slides frictionally over a surface, we balance this elastic force with a friction term, $\Gamma \partial_t u_i$, where $\Gamma$ is the friction coefficient. Such a system can be described by a two-dimensional model with the equation of motion:
\begin{equation}
\label{model2D}
\Gamma \partial_t u_i = \frac{E_2}{2(1+\nu_2)} \left[ \partial_j \partial_j u_i + \frac{1}{1-\nu_2} \partial_i \partial_j u_j \right].
\end{equation}
Note that Eq.~\ref{model2D}, which describes the response of elastic medium to a displacement $u_i$, is similar to the diffusion equation, but is a tensor equation instead of a scalar one.

The second model that we consider is a three-dimensional realization of an overdamped elastic medium, such as a polymer network immersed in a fluid.  The elasticity of the system is also described by Eq.~\ref{stressstrain3D}, but the friction force is proportional to the relative motion of the fluid and the elastic network: $\Gamma( \partial_t u_i - v_i)$. The stress in an incompressible fluid depends linearly on the pressure $p$ and the shear rate, $\dot \gamma_{ij}^\mathrm{visc} = \frac12 ( \partial_i v_j + \partial_j v_i)$:
\begin{equation}
\label{fluidstressstrain}
\sigma_{ij}^\mathrm{visc} = - p \delta_{ij} + 2 \eta \dot \gamma_{ij}^\mathrm{visc},
\end{equation}
in both two and three dimensions~\cite{landaufluidmechanics}. In the over damped limit (zero Reynolds number), taking the divergence of~(\ref{fluidstressstrain}) gives the Stokes equation. Combining the elastic and fluid equations gives a closed system for $u_i$, $v_i$ and $p$:
\begin{eqnarray}
\label{ueq}
\Gamma(\partial_t u_i - v_i) &=& \frac{E}{2(1+\nu)} \left[ \partial_j \partial_j u_i + \frac{1}{1-(d-1)\nu} \partial_i \partial_k u_k \right], \\
\label{veq}
\Gamma(\partial_t u_i - v_i) &=& \partial_i p - \eta \partial_j \partial_j v_i, \\
\label{incomp}
0 &=& \partial_j v_j.
\end{eqnarray}
Eqs.~\ref{ueq}-\ref{incomp} are identical to the two-fluid model studied by Levine and Lubensky~\cite{levine01}, but without the inertial terms. 

We now add mechanical excitability as follows. We consider a collection of nodes at positions $\{ \vec R_n \}$, where $n$ indexes the nodes. A node can be activated if some measure of the stress (for example, the absolute value of its largest eigenvalue) exceeds a threshold value~$\alpha$.  If this occurs at time $t$, the node releases additional stress over a time interval $\Delta t$.  For a node at $\vec R_n$ activated at time $t=t_n$, we therefore introduce an extra force into equation~(\ref{model2D}), of the form
\begin{equation}
\label{activeforce}
P^\mathrm{active}_i = \partial_j Q_{ij} \delta(\vec x - \vec R_n) \Theta(t-t_n) \Theta(t_n + \Delta t - t),
\end{equation}
where $Q_{ij}$ is a tensor of rank 2, which in general can have three contributions: a hydrostatic expansion/contraction, a rotation, and a traceless dipole.

We solve for the response of the two-dimensional overdamped elastic medium of Eq.~\ref{model2D} to the active force in Eq.~\ref{activeforce} by deriving the Green's tensor~$G_{ijk} (\vec x, t)$, which relates the displacement $u_k(\vec x, t)$ to a source term $Q_{ij} \delta(\vec x) \Theta(t)$ at the origin at time $t=0$. We find that the material parameters $E_2$, $\nu_2$ and $\Gamma$ combine in two quantities with the dimensions of diffusion constants,
\begin{equation}
\label{diffconsts}
D_1 = \frac{E_2}{(1-\nu_2^2)\Gamma} = \frac{2}{1-\nu_2} \frac{\mu}{\Gamma}, \quad D_2 = \frac{E_2}{2(1+\nu_2) \Gamma} = \frac{\mu}{\Gamma},
\end{equation}
which correspond to motion in the longitudinal and transverse directions respectively, and together completely determine the solution. Here $\mu = E_2/2(1+\nu_2) = E/2(1+\nu)$ is the material's shear modulus, which is the same in two and three dimensions. The resulting Green's tensor is given by:
\begin{eqnarray}
\label{GreenTensor2D}
G_{ijk} (\vec x, t) &=& -\frac{1}{\mu x} \left\{ \left[ \left( \frac{1-\nu_2}{2}+\frac{8 D_2 t}{x^2} \right) e^{-x^2/4D_1 t} - \left(1+\frac{8 D_2 t}{x^2} \right)  e^{-x^2/4D_2 t} \right] \frac{x_i x_j x_k}{x^3} \right. \nonumber\\
&& \left. - \frac{2 D_2 t}{x^2} \left[ e^{-x^2/4D_1 t} -  e^{-x^2/4D_2 t} \right] \phi_{ijk}  + e^{-x^2/4D_2 t} \delta_{ik} \frac{x_j}{x} \right\},
\end{eqnarray}
where  $x = \sqrt{\vec x \cdot \vec x}$ and $\phi_{ijk} = \delta_{ij} \frac{x_k}{x} +\delta_{ik} \frac{x_j}{x} +\delta_{jk} \frac{x_i}{x}$.

We can derive a similar solution for the response of the two-fluid model of Eqs.~\ref{ueq}-\ref{incomp} to the active force in Eq.~\ref{activeforce}. In this case there is an extra parameter, the viscosity $\eta$ of the fluid, which gives rise to a natural relaxation timescale $\tau = \mu/\eta$ of the system. In three dimensions, the three quantities governing the solution of equations~(\ref{ueq}-\ref{incomp}) are given by:
\begin{eqnarray}
\label{D13D}
D_1 &=& \frac{E}{\Gamma} \frac{1-\nu}{(1+\nu)(1-2\nu)} = \frac{2(1-\nu)}{1-2\nu} \frac{\mu}{\Gamma},\\
\label{D23D}
D_2 &=& \frac{E}{\Gamma} \frac{1}{2(1+\nu)} = \frac{\mu}{\Gamma},\\
\label{tau}
\tau &=& \frac{E}{2\eta(1+\nu)} = \frac{\mu}{\eta}.
\end{eqnarray}
For the associated Green's tensor we find:
\begin{eqnarray}
\label{Gijk3D}
G_{ijk}(\vec x, t) &=& G_{ijk}^\mathrm{hom}(\vec x, t) + G_{ijk}^\mathrm{stat}(\vec x, t), \\
\label{Gijkhom3D}
G^\mathrm{hom}_{ijk}(\vec x, t) &=& -\frac{1}{(2\pi x)^2 D_1 \Gamma} \left[ A\left(\frac{D_1 t}{x^2}\right) \frac{x_i x_j x_k}{x^3} - B\left(\frac{D_1 t}{x^2}\right) \phi_{ijk} \right] \nonumber\\
&& + \frac{e^{-t/\tau}}{(2\pi x)^2 D_2 \Gamma} \left[ A\left(\frac{D_2 t}{x^2}\right) \frac{x_i x_j x_k}{x^3} - B\left(\frac{D_2 t}{x^2}\right) \phi_{ijk} \right] \nonumber\\
&& -\frac{e^{-t/\tau}}{(2\pi x)^2 D_2 \Gamma} C\left(\frac{D_2 t}{x^2}\right) \frac{x_j}{x} \delta_{ik}, \\
\label{Gijkstat3D}
G^\mathrm{stat}_{ijk}(\vec x, t) &=& -\frac{2\pi(1+\nu)}{(2 \pi x)^2E} \delta_{ik} \frac{x_j}{x} - \frac{3 (1+\nu)}{8 \pi x^2 E (1-\nu)} \frac{x_i x_j x_k}{x^3} + \frac{(1+\nu)}{8 \pi x^2 E (1-\nu)} \phi_{ijk},
\end{eqnarray}
where
\begin{eqnarray}
\label{defA}
A(y) &=& \left(15 \sqrt{\pi y} + \sqrt{\frac{\pi}{y}} \right) e^{-1/4y} + \left( \frac32 - 15 y \right) \pi \erf\left(\frac{1}{2\sqrt{y}}\right), \\
\label{defB}
B(y) &=& 3 \sqrt{\pi y} e^{-1/4y} + \left( \frac12 - 3 y \right) \pi \erf\left(\frac{1}{2 \sqrt{y}}\right), \\
\label{defC}
C(y) &=&  \sqrt{\frac{\pi}{y}} e^{-1/4y} - \pi \erf\left(\frac{1}{2\sqrt{y}} \right).
\end{eqnarray}
For an input term that runs only over a time interval $\Delta t$ as in Eq.~\ref{activeforce}, continuity demands that for $t<\Delta t$ we have $G_{ijk}(\vec x, t) = G^\mathrm{hom}_{ijk}(\vec x, t) + G^\mathrm{stat}_{ijk}(\vec x, t)$, and for $t>\Delta t$ this changes to $G_{ijk}(\vec x, t) = G^\mathrm{hom}_{ijk}(\vec x, t) - G^\mathrm{hom}_{ijk}(\vec x, t-\Delta t)$. 

\begin{figure}[tb]
\begin{center}
\begin{tabular}{c}
\includegraphics[scale=2]{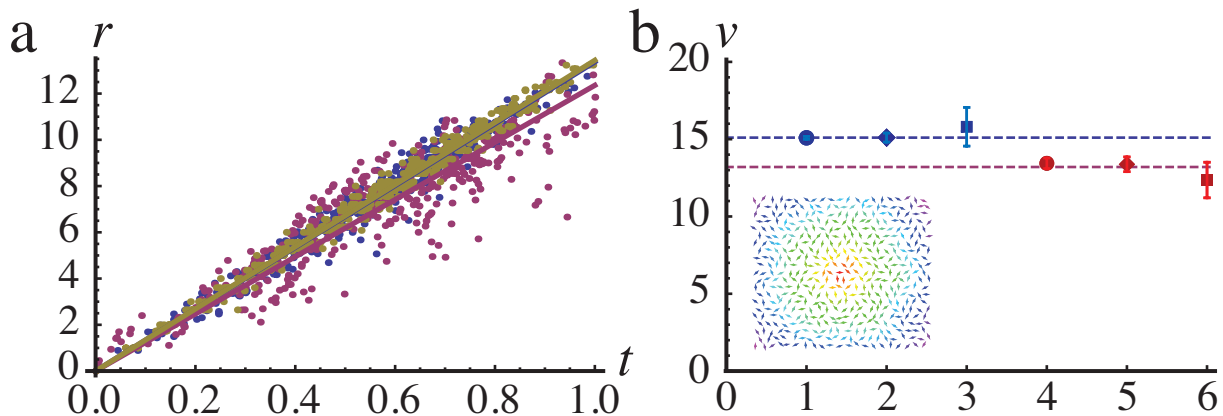}
\end{tabular}
\end{center}
\caption{Calculated wavefronts. (a) Plots showing for each node (dots) the distance from the first activated node vs. time of activation, with linear fits. This example has a dipole source term and shows results for three different types of grids, all with the same density: regular triangular (gold), correlated random (blue) and uncorrelated random (red). (b) Mean wavefront speed for six realizations, with hydrostatic force term (1-3, blue) and dipole force terms (4-6, red), and on three different grids: regular triangular (circle, 1 \& 4), correlated random (diamond, 2 \& 5), and uncorrelated random (square, 3 \& 6). Error bars indicate standard deviations. Inset shows the 2-dimensional field of nodes, indicating the orientation in which nodes are activated, and color-coded according to the time at which they are activated, on a hue scale (red-yellow-green-blue-violet).}
\label{fig:mechanicalwavefront}
\end{figure}

Because our model equations are linear, we can now use the principle of superposition to study the effect of many source terms. We initialize the system by activating a single node at the origin at $t=0$. We then measure the stress at the other nodes as a function of time, and activate them if they are above threshold by releasing more stress, according to equation~(\ref{activeforce}). We consider various cases for the arrangement of the nodes: a regular triangular lattice, a random configuration with short-range correlations (as in a random packing of disks) and an uncorrelated random configuration. In addition, we look at variants in which the force term is purely isotropic (hydrostatic expansion/contraction) or is in the form of a volume-conserving force dipole, with either random orientation or orientations correlated to the direction of the traveling wavefront. In all cases, the model produces an activation wavefront with a well-defined speed, as shown in Fig.~\ref{fig:mechanicalwavefront}.  We find that the speed of the wavefront depends on the density of nodes but is insensitive to their arrangement.  However, the spread of the wavefront around its mean increases with the amount of randomness (Fig.~\ref{fig:mechanicalwavefront}(b)). Not surprisingly, if the orientations of the force dipoles are chosen at random, the speed of the wavefront is the same in all directions so that its shape is circular, as in Fig.~\ref{fig:mechanicalwavefront}(b, inset). In contrast, if the dipoles are all oriented in the same direction, the wavefront is no longer uniform, but is faster in the direction of orientation. Also, for the same magnitude of the active force, the speed of the wavefront is somewhat higher if the force dipole $Q_{ij}$ is hydrostatic than if it is a traceless dipole (Fig.~\ref{fig:mechanicalwavefront}(b)). All these observations indicate that the wavefront speed is dictated primarily by the properties of the medium and the average distance between nodes, and is insensitive to both the form of the active force and spatial arrangement of nodes.

\begin{figure}[tb]
\begin{center}
\begin{tabular}{c}
\includegraphics[scale=2]{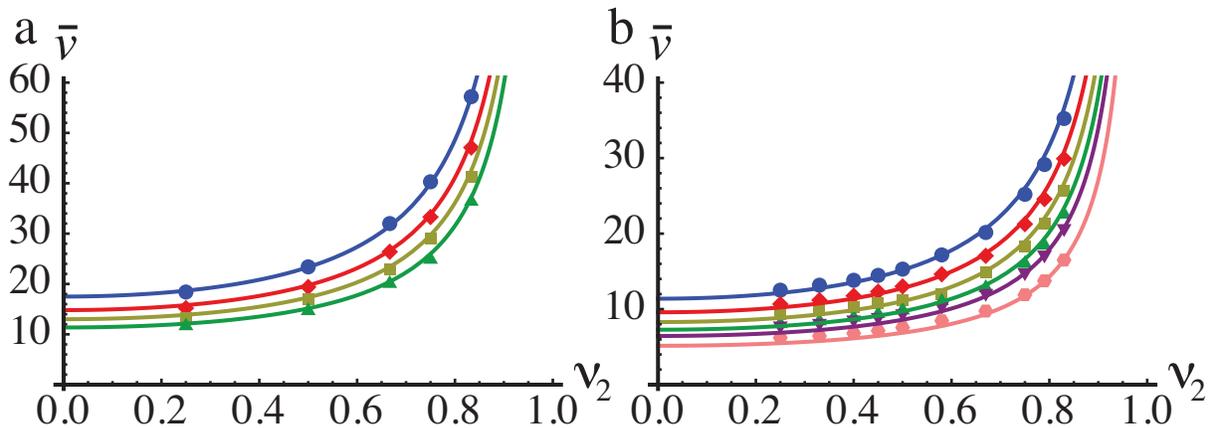}
\end{tabular}
\end{center}
\caption{Dimensionless wavefront speed as a function of the Poisson ratio $\nu_2$ and dimensionless threshold $\bar{\alpha}$. Symbols indicate numerical solutions of the full system, lines the functional form of equation~(\ref{wavefrontspeed}). (a) Hydrostatic force term $Q_{ij} = Q \delta_{ij}$. Fit parameters $c_1 = 4.0$, $c_2 = 1.5$. Values of $\bar{\alpha}$: 0.1 (blue/dots), 0.2 (red/ diamonds), 0.3 (gold/squares) and 0.4 (green/triangles). (b) Dipole force term with random orientation angle~$\theta$: $Q_{ij} = -Q \cos(2\theta) (\delta_{i1}\delta_{j1} - \delta_{i2}\delta_{j2}) - Q \sin(2\theta) (\delta_{i1}\delta_{j2} + \delta_{i2}\delta_{j1})$. Fit parameters $c_1 = -1.0$, $c_2 = 1.0$. Values of $\bar{\alpha}$: 0.050 (blue/dots), 0.075 (red/diamonds), 0.100 (gold/squares), 0.125 (green/triangles up), 0.150 (purple/triangles down), and 0.200 (pink/hexagons).}
\label{fig:wavefrontspeed}
\end{figure}

Dimensional analysis of the material parameters of our two-dimensional model~(\ref{model2D}) shows that there is only one possible scaling for the wavefront speed with the material parameters: $v \sim E_2 / (a \Gamma)$, where $a$ is the grid spacing. The dimensionless speed $\bar{v} = (a \Gamma/E_2) v$ depends on the material's Poisson ratio~$\nu_2$ and the dimensionless threshold $\bar{\alpha} = \alpha*a^2/Q$, where $Q$ is the strength of the force term. We have determined the function $\bar{v}(\nu_2, \bar{\alpha}$ numerically for both hydrostatic and dipole force terms. We find that it obeys a fairly simple functional form, which can be motivated by an analytical argument based on the case of the simplest force term, the purely hydrostatic one $Q_{ij} = Q \delta_{ij}$. In this case, the stress is given by
\begin{equation}
\label{hydrostaticstress}
\sigma_{kl}^\mathrm{hydro} = \frac{(1-\nu_2)Q}{x^2} \left[ \left( 2 + \frac{x^2}{2 D_1 t} \right) \frac{x_k x_l}{x^2} + \left( \frac{\nu_2}{1-\nu_2} \frac{x^2}{2 D_1 t} - 1 \right) \delta_{kl} \right] e^{-x^2/4 D_1 t}.
\end{equation}
Because the stress drops off quadratically with distance, the major contribution to the stress at any node is due to forces exerted by neighboring nodes. Moreover, since the front expands radially, typically only a single nearest neighbor of any node will have been activated recently. We can therefore get a reasonable estimate for the local stress at a node by considering that nearest neighbor to be the only source. We introduce the dimensionless time $\bar{t} = E_2 t / (a^2 \Gamma)$; then for a single source a distance~$a$ away, the time at which the largest eigenvalue of the stress~(\ref{hydrostaticstress}) reaches the dimensionless threshold $\bar{\alpha}$ is given by:
\begin{equation}
\label{wavefronttime}
\bar{\alpha} = (1-\nu_2) \left[ 1 + \frac{1+\nu_2}{2 \bar{t}} \right] e^{-(1-\nu_2^2)/4\bar{t}}.
\end{equation}
Unfortunately, Eq.~(\ref{wavefronttime}) cannot be inverted analytically. However, the two factors containing $\bar{t}$ are easily inverted, allowing us to make an educated guess for the functional form of the resulting dimensionless speed:
\begin{equation}
\label{wavefrontspeed}
\bar{v} = -\frac{4(c_1 \bar{\alpha} + c_2) \log(\bar{\alpha})}{1-\nu_2^2},
\end{equation}
where $c_1$ and $c_2$ need to be determined numerically; we find $c_1 = 4.0$ and $c_2 = 1.5$. As shown in Fig.~\ref{fig:wavefrontspeed}(a), the form given by equation~(\ref{wavefrontspeed}) works remarkably well. Moreover, the same functional form also describes the results for a dipole force term wavefront, as shown in Fig.~\ref{fig:wavefrontspeed}(b), the only difference being the values of the two fit parameters - here we find $c_1 = -1.0$ and $c_2 = 1.0$.

In line with intuition, our model predicts that there is a maximum threshold value $\bar{\alpha}_\mathrm{max}$ above which a wavefront will not propagate. This can happen for one of two reasons: either the force is not large enough to create a stress at the next node that exceeds the threshold value, or the nodes are so far apart that, due to the diffusive nature of the stress spreading, the threshold value is not reached. Both possibilities are contained in the form of the dimensionless version of $\bar{\alpha}_\mathrm{max}$, given by the maximum of the right hand side of equation~(\ref{wavefronttime}), which gives $\bar{\alpha}_\mathrm{max} = 2 e^{-(1+\nu_2)/2}$ at $\bar{t}_\mathrm{max} = \frac{1-\nu_2}{2}$ and corresponding to a minimum speed $\bar{v}_\mathrm{min} = \frac{2}{1-\nu_2}$. We note that as $\nu_2$ approaches its maximum value of 1, the minimum speed diverges, as can be seen in Fig.~\ref{fig:wavefrontspeed}(a). 

For the three-dimensional two-fluid model of Eqs.~(\ref{ueq}-\ref{incomp}), there are two independent quantities with the dimensions of speed,  $E/a\Gamma$ and $a/\tau$, where $\tau = \eta/\mu$ is the material's relaxation time (Eq.~\ref{tau}). We note that both of these scale linearly with the material's Young's modulus~$E$ (or equivalently, with the material's shear modulus~$\mu$), which implies that also in this case the resulting wavefront velocity in a similar setup with excitable nodes will scale linearly with that modulus. It will also scale with $\Gamma^{-n} \eta^{1-n}$, where $n$ is some number between 0 and 1, indicating that both the internal viscosity of the moving fluid and the friction between the elastic and viscous material contribute to the damping of the ballistic motion. 

In this paper, we have introduced two theoretical realizations of mechanical signaling in an overdamped elastic medium.  We have shown that nonlinear wavefront propagation in the models is a robust feature of both models.  In both cases, the wavefront velocity is insensitive to the spatial distribution of excitable nodes.  It is also insensitive to whether the stress is released in an isotropic or traceless anisotropic fashion.  Furthermore, a fundamental feature of both models is that the wavefront velocity is proportional to the Young's modulus of the medium, and the magnitude of the velocity can be understood simply and quantitatively in terms of characteristic dimensionless variables, such as the stress threshold made dimensionless with the magnitude of the force dipole released when a node is excited. 

The overdamped elastic models considered here are the simplest models that could be used to describe a tissue. It would be worthwhile to explore mechanical signaling in other models that have been proposed for tissues, including the active gel model~\cite{bois11,kruse04,kruse05} and cellular models~\cite{manning10,chiou12}. The active gel model of Kruse \textit{et al.}~\cite{kruse04,kruse05} is an extension of the two-fluid model of Levine and Lubensky~\cite{levine01} with a continually active (energy-consuming) term to model the dynamics of the cytoskeleton due to motor activity. As we have shown here, such a continuous activity is not necessary to describe wavefront propagation, as local and discrete activity is sufficient. However, given the presence of active motors in the cytoskeleton, it would be interesting to see how the wavefront is affected by an active term in the model. It would also be interesting to compare the results of such an interaction with those of Bois \textit{et al.} who study pattern formation in active fluids due to chemical signaling~\cite{bois11}. These active models, and the one we used here, are continuous models. However, tissues are of course composed of discrete units, the cells. As shown by Manning \textit{et al.} several mechanical properties of the tissue, like its surface tension, are determined by cell-cell adhesion and cortical tension~\cite{manning10}. Recent work by Chiou \textit{et al.} provides a method to measure the relative magnitude of forces acting within and between cells~\cite{chiou12}. These results now make it possible to construct a quantitative cell-based tissue model, in which wavefront propagation due to mechanical signaling can be studied as well.

Now that we have introduced a minimal model for mechanical signaling via nonlinear wavefront propagation, we can ask how one might identify biological contexts in which mechanical signaling is likely to occur.  We can also ask how to determine whether a given wavefront is an example of mechanical signaling.  Wavefronts of processes that generate stresses are obvious likely candidates.  In order for a medium to be mechanically-excitable, however, it is not enough to have a collection of nodes capable of generating stress.  There must also be a mechanism to trigger the nodes to release stress once a stress threshold is reached.  A likely mechanism would involve stress-dependent ion channels that can release ions above a threshold stress~\cite{gillespie01,kung05,kung10,sukharev12}. Thus, a system with known wavefronts of ion concentration would be a possible candidate, especially if the wavefronts can also be triggered mechanically.  Mitotic wavefronts in the early {\it Drosophila} embryo fit many elements of this profile~\cite{idema13plos}.  The process of mitosis generates stresses as chromosomes condense and as they segregate~\cite{kleckner04}. There is a known calcium wavefront that propagates across the embryo in tandem with the mitotic wavefront~\cite{parry05}.  Similarly, contractile wavefronts in heart tissue may be a form of mechanical signaling~\cite{majkut13}; they generate stresses as cardiomyocytes contract in a process that is well known to involve calcium via the excitation-contraction mechanism~\cite{huxley69,ashley91}. Another possible realization is spreading cortical depression, which involves a potassium wavefront~\cite{gorelova83} and that can be triggered mechanically~\cite{akerman08}.  These examples suggest that it may be worthwhile to re-examine other known examples of ion signaling in biological contexts to see if they are more properly interpreted as mechanical signaling.

We thank Gareth Alexander, Michael Lampson, Tom Lubensky and Phil Nelson for instructive discussions. This work was partially supported by the Netherlands Organization for Scientific Research through a Rubicon grant (T.I.) and by NSF-DMR-1104637 (A.J.L.).


\begin{thebibliography}{20}
\bibitem{devreotes89}
P. Devreotes, Science {\bf 245}, 1054 (1989).

\bibitem{levine91}
H. Levine and W. Reynolds, Phys. Rev. Lett. {\bf 66}, 2400 (1991).

\bibitem{lee96}
K. J. Lee, E. C. Cox, and R. E. Goldstein, Phys. Rev. Lett. {\bf 76}, 1174 (1996).

\bibitem{lechleiter92}
J. D. Lechleiter and D. E. Clapham, Cell {\bf 69}, 283 (1992).

\bibitem{gorelova83}
N. A. Goroleva and J. Bures, J. Neurobiol. {\bf 74}, 353 (1983).

\bibitem{altman02}
G. H. Altman, \textit{et al.}, FASEB J. {\bf 16} 270 (2002).

\bibitem{discher05}
D. E. Discher, P. Janmey, and Y-L. Wang, Science {\bf 310}, 1139 (2005).

\bibitem{vogel06}
V. Vogel and M. Sheetz, Nat. Rev. Mol. Cell Biol. {\bf 7}, 265 (2006).

\bibitem{discher09}
D. E. Discher, D. J. Mooney, and P. W. Zandstra, Science {\bf 324}, 1673 (2009).

\bibitem{subramony13}
S. D. Subramony \textit{et al.}, Biomaterials {\bf 34}, 1942 (2013).

\bibitem{bois11}
J. S. Bois, F. J\"ulicher, and S. W. Grill, Phys. Rev. Lett. {\bf 106}, 028103 (2011).

\bibitem{serra12}
X. Serra-Picamal \textit{et al.}, Nature Phys. {\bf 8}, 628 (2012).

\bibitem{idema13plos}
T. Idema, J. O. Dubuis, M. L. Manning, P. C. Nelson and A. J. Liu, submitted.

\bibitem{majkut13}
S. Majkut \textit{et al.}, submitted to Curr. Biol.

\bibitem{foe83}
V. E. Foe and B. M. Alberts, J. Cell Sci. {\bf 61}, 31 (1983).

\bibitem{dejong}
F. de Jong \textit{et al.}, Circ. Res. {\bf 71}, 240 (1992).

\bibitem{landauelasticity}
L.~D. Landau and E.~M. Lifshitz, \textit{Theory of Elasticity}, 3rd ed. (Butterworth Heinemann, Burlington, MA, U.S.A., 1986).

\bibitem{landaufluidmechanics}
L.~D. Landau and E.~M. Lifshitz, \textit{Fluid Mechanics}, 2nd ed. (Butterworth Heinemann, Burlington, MA, U.S.A., 1987).

\bibitem{levine01}
A.~J. Levine and T.~C. Lubensky, Phys. Rev. E {\bf 63},  041510  (2001).

\bibitem{kruse04}
K. Kruse, J.-F. Joanny, F. J\"ulicher, J. Prost, and K. Sekimoto, Phys. Rev. Lett. {\bf 92}, 078101 (2004).

\bibitem{kruse05}
K. Kruse, J.-F. Joanny, F. J\"ulicher, J. Prost, and K. Sekimoto, Eur. Phys. J. E {\bf 16}, 5 (2005).

\bibitem{manning10}
M. L. Manning, R. A. Foty, M. S. Steinberg, and E.-M. Schoetz, Proc. Nat. Acad. Sci. USA {\bf 107}, 12517 (2010).

\bibitem{chiou12}
K. K. Chiou, L. Hufnagel, and B. I. Shraiman, PLoS Comput. Biol. {\bf 8}, e1002512 (2012).

\bibitem{gillespie01}
P. G. Gillespie and R. G. Walker, Nature {\bf 413}, 194 (2001).

\bibitem{kung05}
C. Kung, Nature {\bf 436}, 647 (2005).

\bibitem{kung10}
C. Kung, B. Martinac, and S. Sukharev, Annu. Rev. Microbiol. {\bf 64}, 313 (2010).

\bibitem{sukharev12}
S. Sukharev and F. Sachs, J. Cell Sci. {\bf 125}, 3075 (2012).

\bibitem{kleckner04}
N. Kleckner \textit{et al.}, Proc. Natl. Acad. Sci. USA {\bf 101}, 12592 (2004).

\bibitem{parry05}
H. Parry, A. McDougall and M. Whitaker, J. Cell Biol. {\bf 171}, 47 (2005).

\bibitem{huxley69}
H. E. Huxley, Science {\bf 164}, 1356 (1969).

\bibitem{ashley91}
C. C. Ashley, I. P. Mulligan, and T. J. Lea, Quarterly Rev. Biophys. {\bf 24}, 1 (1991).

\bibitem{akerman08}
S. Akerman, P. R. Holland and P. J. Goadsby, Brain Res. {\bf 1229}, 27 (2008).

\end{thebibliography}
\end{document}